\documentclass[aps,prd,preprint,onecolumn,notitlepage,eqsecnum,showkeys]{revtex4-2}

\usepackage{hyperref}

\def\Q{\overrightarrow{Q}}
\renewcommand{\P}{\overrightarrow{P}}

\newcommand{\pp}{\overrightarrow{p}}

\newcommand{\qq}{\overrightarrow{q}}

\def\half{\textstyle{\frac{1}{2}}}

\def\ra{\rightarrow}

\def\d{\partial}

\def\b{\begin{eqnarray*}}  %takes no eqn numbers
\def\e{\end{eqnarray*}}    %takes no eqn numbers
\def\bn{\begin{eqnarray}}  %takes eqn numbers
\def\en{\end{eqnarray}}   %takes eqn numbers

\def\<{\langle}
\def\>{\rangle}

\def\{{\lbrace}

\def\d3{d^3\!x}

\def\b{\beta}

\def\}{\rbrace}
\begin{document}     

\title{Thank The Quantum Realm For {\it Nothing} Ever Entering Into Black Holes}
\author{John R. Klauder}
\email{klauder@ufl.edu}
\affiliation{Department of Physics and Department of Mathematics, 
University of Florida, Gainesville, USA}
\author{Riccardo Fantoni}
\email{riccardo.fantoni@scuola.istruzione.it}
\affiliation{Universit\`a di Trieste, Dipartimento di Fisica, strada
  Costiera 11, 34151 Grignano (Trieste), Italy}

\date{\today} 

\begin{abstract} While the quantum realm seems hidden, it can also reach examples of infinite energy, especially when a part of space is roughly removed until it disappears, possibly forever. Since it follows that {\it Nothing } can enter a region where the space is missing, the quantum realm, as seen now in affine quantization,  will automatically come to help everything else by creating colossal `quantum walls' that will ensure that everything stays out of all  black holes.

In this article, we show that the expanded quantum realm allows {\it Nothing} to ever fall into a black hole.
\end{abstract} 


\maketitle 
 
\section{Introduction} 

While humanity believes that the quantum realm is very weak due to a very small $\hbar$-factor, they do not appreciate that it can be immensely strong because, effectively, selected  $\hbar$-terms can reach  an infinite energy.

\section{A Brief Examination of the Current Understanding of Black Holes}

The current understanding of black holes is naturally from a classical viewpoint, and, to a degree,  also relies  on  canonical quantization. In this case it requires  that the physical space  exists everywhere, e.g., $-\infty<q<\infty$. Black holes  have then, essentially, found that colossal quantities of trash  have piled up, which  are so huge that they  can even develop very strong attractions of gravity. 

\section{A  New and Very Different Analysis }

\subsection{A new quantization procedure} 

First of all, we introduce a new, and different, formulation of quantization procedures, which is known as  {\bf Affine Quantization}. It has the very important property that it can  only exist  if there are   {\it  missing parts of space}, and that is shown by incomplete regions that  still  remain, e.g., $0<q$, $0<|q|$,  $|q|<b$, and especially   $|q|>b>0$, since the last example will be important for this article, and which could offer a very great distinction between this new understanding of black holes and the  conventional understanding of black holes. This new quantization  procedure has proven itself very well in dealing with  examples that   only have an   {\bf incomplete space}. 

Now we will examine what could look like regions of missing space that  might lead us to investigate and  see if that might be black holes.  
 
\subsection{A toy model that is VERY relevant}

The  classical Hamiltonian for the half-harmonic oscillator is
$H=(p^2+q^2)/2$, but it has been so-named  because  we   have  choosen  that  only $q>0$  remains. 	In that case, and using affine quantization, we find that the quantum Hamiltonian is ${\cal{H}}=(P^2+(3/4)\hbar^2/Q^2+Q^2)/2$, with $Q>0$. For this example, the  eigenvalues are $E'_n=2\hbar(n+1)$ for $n=0,1,2,...$, while the eigenvalues for the well-known  full-harmonic oscillator are  $E_n=\hbar(n+1/2)$. Evidently, in each case, the eigenvalues are equally spaced, and the number 2   `has just played an important role' \cite{Gouba2020}.

Since the $\hbar$-term can become  very, very, strong close to $q=0$, it would be useful to introduce  a new way  of spelling the word classical, namely by  {\bf classicAL},  as a {\it  signal} that all  $\hbar$-terms have been {\bf included} along with  the standard  classical elements  used for standard  classical equations. Specifically, we would now  like to use $H=(p^2+(3/4)\hbar^2/q^2+q^2)/2$,  because now  $q>0$. In fact, that would help signal that its classical particles must bounce backwards at the point $q=0$. Since each of these potential-like $\hbar$-terms can even reach infinity,  it seems only reasonable that  
such $\hbar$-terms should  appear together with  standard polynomials in the same kind of equations. After all,  you would  readily accept  $ H=(p^2+\gamma 10^{-50}q^{-50} +q^2)/2$ into the classical Hamiltonian family, so why not let suitable   $\hbar$-terms that {\it could, should, and would}, act as very useful potentials since they can reach infinity.  In addition, 
it is also noteworthy that when  affine quantization immediately finds a new missing  space it  {\bf automatically} introduces  a new  `quantum wall' to keep everything away from that missing space.  

 In addition, if  the remaining  spatial space, $q>0$, was partially increased, by setting it now be   $(q+b)>0$, with $b>0$, then the new eigenvalues would  still be equally spaced, and  finally, when  $b\rightarrow \infty $, we would correctly obtain all of the properties  of   its canonical quantization \cite{Handy2021a}. Effectively,  all  affine  expressions, can  eventually  become a  related canonical formulation  just by restoring all of the missing space. 

 Returning to our initial expressions, observe that $q>0$, which is our retained space, and refusing from entering our missing space, which, for this example, is $q\leq 0$, even having a classical Hamiltonian, such as $H=(p^2+q^2)/2$, when $q\leq 0$ is  no longer prohibited. It would seem to make  more  useful physics by adding the  $\hbar$-term that was `smart enough' to  become  this kind of classicAL Hamiltonian,  e.g., $H=(p^2+(3/4)\hbar^2/q^2+q^2)/2$.

Surprisingly, our study of black holes will not be so very different than the topics of this section. 

{\bf  Remember that $\hbar^2$ is NOT ZERO, and that affine quantization can  rigorously, and correctly, solve a very different set of problems  than those of  canonical  quantization: specifically,   affine quantization has been  designed to solve   ALL examples with ANY kind of missing  space. } 

\subsection{Exploiting missing parts of space} 

We shall propose  that black holes may  be examined through a  specific, and correct, article, namely  {\bf The Particle in a Box Warrants Examination} \cite{Klauder2022}. All that would be necessary, effectively, is to use two, very similar,  Hamiltonians except that the first one is active  {\bf inside} the box, while  the second one is active {\bf outside} the box. In order to  deal with the one being outside, it will be necessary  to  add additional potentials that can handle the complete  outside space. Effectively, the second example space just has  a finite section  removed from  the complete space. 
The Hamiltonians for the `particle in a box', have been correctly created in the article just mentioned.
\footnote{The details of the $\hbar$-terms have been taken from a special article, and while they can reach the 3/4 story extremely close to each end, the analysis in between the two ends  requires special attention from page 3 in \cite{Fantoni22b}.}

It is physically correct that the $\hbar$-terms should now also belong with the classicAL  family, which has a `new spelling and  meaning', that signals that this word is  now being used  to  {\bf include all}   $\hbar$-terms as well,  especially because    all   
of those   $\hbar$-terms  can already reach infinity,  and should be allowed alongside the conventional  potential terms  in standard classical Hamiltonian equations. Now, both the classicAL and quantum Hamiltonians have  included all $\hbar$-terms,  such as in this example, first offered in the new classicAL form, 
\bn H= \half[p^2/m+ \hbar^2(2q^2+b^2)/(q^2-b^2)^2  +m \;q^2] +V(q),
\label{1}\en
 and second,   and using the standard   Schr\"odinger   formulation,  leads to  
\bn {\cal{H}}= \half[P^2/m+ \hbar^2(2q^2+b^2)/(q^2-b^2)^2 +m \;q^2]+ V(q). \label{2} \en

To make this example even more physical-like, we can just let $b\ra b(x,t)$ in those two equations. Also, 2 or 3 spatial 
dimensions can be accepted, simply by changimg
$p\ra  \pp$ and $q\ra \qq$  as well as
$P\ra \P$ and $Q\ra \Q$.
\footnote{Although  we will not need it in this article, but if the missing space is just a {\it single point} in space,  like just removing $q= 0$, that would still require  `quantum walls', as seen  in equations (1) and (2), the factor 
 $b$ can be reduced to zero indicating that  removing just a single point from space will  still create `quantum walls'. That final expression could represent the birth (or death) of a black hole.}

{\bf The authors believe that strong  $\hbar$-terms definitely belong in  classicAL physics, and be  included  in the traditional classical equations that would be suitable, because terms, such as $2\hbar^2/q^2$, which can reach infinity, should be serving as  standard potentials, and certainly such terms would deserve to be added to appropriate classicAL equations. }

\section{Summary and Outlook} 

 The authors  also believe that  space can be broken  which  leads to regions that are completely absent of any space. Of course, that would have taken  huge physical efforts, but when enough trash has been  closely assembled, it is very likely that it could easily destroy the local space through fantastic weight, giant explosions, and any other brutal crushing.

As the toy model in this article has showed, that after suddenly removing  a specific portion of space, the quantum realm will {\it instantly } create a `quantum wall' in order to  keep everything out of the new missing space.  

Now the  new, and very useful, affine quantization procedures will {\bf  never} let anything fall into such a black hole. In addition, it is noteworthy that you should realize the ability to stop anything from entering a black hole because   the   new quantization process will  keep everything out of any black hole.  

You may  be assured  that canonical quantization may  never  be suitable to deal with black holes. Fortunately,    this new, and very suitable, affine quantization, and its procedures, can definitely deal with incomplete space,  and  AQ is now available to have it be fully explained. If you wish  to have a better understanding of affine  quantization, you could examine a beginner's article \cite{Klauder2020c}, or for more experienced people, choose article \cite{Fantoni23b}.
 
{\bf And finally, black holes are dealing with giant fires that are held surrounding  the black hole itself. It is proposed now that a `quantum wall'  will hold trash away from entering  into the black hole, and, like any other potential would do, it will  glow  from  the burning trash, which has piled up against the  `quantum wall' surrounding  closely just outside of the black hole. The conventional view may not have as good an understanding of the `ring of fire' surrounding a black hole as does  that of the quantum realm formulation. 

The authors of this article are convinced that the standard description of black holes is very likely incorrect, and since Nature has  definitely showed us how to make a  `quantum wall', the contents of this article should appear more believable than those of the standard  understanding of black holes.}

\subsection*{\bf ACKNOWLEDGMENTS}
The authors wish to thank Dr. Dustin Wheeler for assistance in typesetting and editing the article.

\bibliography{killer-3p}

\begin{thebibliography}{8}%
\makeatletter
\providecommand \@ifxundefined [1]{%
 \@ifx{#1\undefined}
}%
\providecommand \@ifnum [1]{%
 \ifnum #1\expandafter \@firstoftwo
 \else \expandafter \@secondoftwo
 \fi
}%
\providecommand \@ifx [1]{%
 \ifx #1\expandafter \@firstoftwo
 \else \expandafter \@secondoftwo
 \fi
}%
\providecommand \natexlab [1]{#1}%
\providecommand \enquote  [1]{``#1''}%
\providecommand \bibnamefont  [1]{#1}%
\providecommand \bibfnamefont [1]{#1}%
\providecommand \citenamefont [1]{#1}%
\providecommand \href@noop [0]{\@secondoftwo}%
\providecommand \href [0]{\begingroup \@sanitize@url \@href}%
\providecommand \@href[1]{\@@startlink{#1}\@@href}%
\providecommand \@@href[1]{\endgroup#1\@@endlink}%
\providecommand \@sanitize@url [0]{\catcode `\\12\catcode `\$12\catcode
  `\&12\catcode `\#12\catcode `\^12\catcode `\_12\catcode `\%12\relax}%
\providecommand \@@startlink[1]{}%
\providecommand \@@endlink[0]{}%
\providecommand \url  [0]{\begingroup\@sanitize@url \@url }%
\providecommand \@url [1]{\endgroup\@href {#1}{\urlprefix }}%
\providecommand \urlprefix  [0]{URL }%
\providecommand \Eprint [0]{\href }%
\providecommand \doibase [0]{https://doi.org/}%
\providecommand \selectlanguage [0]{\@gobble}%
\providecommand \bibinfo  [0]{\@secondoftwo}%
\providecommand \bibfield  [0]{\@secondoftwo}%
\providecommand \translation [1]{[#1]}%
\providecommand \BibitemOpen [0]{}%
\providecommand \bibitemStop [0]{}%
\providecommand \bibitemNoStop [0]{.\EOS\space}%
\providecommand \EOS [0]{\spacefactor3000\relax}%
\providecommand \BibitemShut  [1]{\csname bibitem#1\endcsname}%
\let\auto@bib@innerbib\@empty
%</preamble>
\bibitem [{\citenamefont {Gouba}(2021)}]{Gouba2020}%
  \BibitemOpen
  \bibfield  {author} {\bibinfo {author} {\bibfnamefont {L.}~\bibnamefont
  {Gouba}},\ }\bibfield  {title} {\bibinfo {title} {{Affine Quantization on the
  Half Line}},\ }\href {https://doi.org/10.4236/jhepgc.2021.71019} {\bibfield
  {journal} {\bibinfo  {journal} {Journal of High Energy Physics, Gravitation
  and Cosmology}\ }\textbf {\bibinfo {volume} {7}},\ \bibinfo {pages} {352}
  (\bibinfo {year} {2021})}\BibitemShut {NoStop}%
\bibitem [{\citenamefont {Handy}(2021)}]{Handy2021a}%
  \BibitemOpen
  \bibfield  {author} {\bibinfo {author} {\bibfnamefont {C.}~\bibnamefont
  {Handy}},\ }\bibfield  {title} {\bibinfo {title} {{Affine Quantization of the
  Harmonic Oscillator on the Semi-bounded domain $(-b, \infty)$ for $ b:
  0\rightarrow \infty$}},\ }\href@noop {} {\  (\bibinfo {year} {2021})},\
  \bibinfo {note} {arXiv:2111.10700}\BibitemShut {NoStop}%
\bibitem [{\citenamefont {Klauder}(2022)}]{Klauder2022}%
  \BibitemOpen
  \bibfield  {author} {\bibinfo {author} {\bibfnamefont {J.~R.}\ \bibnamefont
  {Klauder}},\ }\bibfield  {title} {\bibinfo {title} {{Particle in a Box
  Warrants an Examination}},\ }\href
  {https://doi.org/10.4236/jhepgc.2022.83043} {\bibfield  {journal} {\bibinfo
  {journal} {Journal of High Energy Physics, Gravitation and Cosmology}\
  }\textbf {\bibinfo {volume} {8}},\ \bibinfo {pages} {623} (\bibinfo {year}
  {2022})}\BibitemShut {NoStop}%
\bibitem [{Note1()}]{Note1}%
  \BibitemOpen
  \bibinfo {note} {The details of the $\protect \hbar $-terms have been taken
  from a special article, and while they can reach the 3/4 story extremely
  close to each end, the analysis in between the two ends requires special
  attention from page 3 in \cite {Fantoni22b}.}\BibitemShut {Stop}%
\bibitem [{Note2()}]{Note2}%
  \BibitemOpen
  \bibinfo {note} {Although we will not need it in this article, but if the
  missing space is just a {\protect \it single point} in space, like just
  removing $q= 0$, that would still require `quantum walls', as seen in
  equations (1) and (2), the factor $b$ can be reduced to zero indicating that
  removing just a single point from space will still create `quantum walls'.
  That final expression could represent the birth (or death) of a black
  hole.}\BibitemShut {Stop}%
\bibitem [{\citenamefont {Klauder}(2020)}]{Klauder2020c}%
  \BibitemOpen
  \bibfield  {author} {\bibinfo {author} {\bibfnamefont {J.~R.}\ \bibnamefont
  {Klauder}},\ }\bibfield  {title} {\bibinfo {title} {{The Benefits of Affine
  Quantization}},\ }\href {https://doi.org/10.4236/jhepgc.2020.62014}
  {\bibfield  {journal} {\bibinfo  {journal} {Journal of High Energy Physics,
  Gravitation and Cosmology}\ }\textbf {\bibinfo {volume} {6}},\ \bibinfo
  {pages} {175} (\bibinfo {year} {2020})}\BibitemShut {NoStop}%
\bibitem [{\citenamefont {Klauder}\ and\ \citenamefont
  {Fantoni}(2023)}]{Fantoni23b}%
  \BibitemOpen
  \bibfield  {author} {\bibinfo {author} {\bibfnamefont {J.~R.}\ \bibnamefont
  {Klauder}}\ and\ \bibinfo {author} {\bibfnamefont {R.}~\bibnamefont
  {Fantoni}},\ }\bibfield  {title} {\bibinfo {title} {{The Magnificent Realm of
  Affine Quantization: valid results for particles, fields, and gravity}},\
  }\href {https://doi.org/10.3390/axioms12100911} {\bibfield  {journal}
  {\bibinfo  {journal} {Axioms}\ }\textbf {\bibinfo {volume} {12}},\ \bibinfo
  {pages} {911} (\bibinfo {year} {2023})}\BibitemShut {NoStop}%
\bibitem [{\citenamefont {Fantoni}\ and\ \citenamefont
  {Klauder}(2022)}]{Fantoni22b}%
  \BibitemOpen
  \bibfield  {author} {\bibinfo {author} {\bibfnamefont {R.}~\bibnamefont
  {Fantoni}}\ and\ \bibinfo {author} {\bibfnamefont {J.~R.}\ \bibnamefont
  {Klauder}},\ }\bibfield  {title} {\bibinfo {title} {{Kinetic Factors in
  Affine Quantization and Their Role in Field Theory Monte Carlo}},\ }\href
  {https://doi.org/10.1142/S0217751X22500944} {\bibfield  {journal} {\bibinfo
  {journal} {Int. J. Mod. Phys. A}\ }\textbf {\bibinfo {volume} {37}},\
  \bibinfo {pages} {2250094} (\bibinfo {year} {2022})}\BibitemShut {NoStop}%
\end{thebibliography}%

\end{document}